\documentstyle[general_cite,psfig]{mn}
\title{Are X-ray properties of loose groups different from those of compact
  groups?}
\author[Stephen F. Helsdon and Trevor J. Ponman]
       {Stephen F. Helsdon\thanks{E-mail: sfh@star.sr.bham.ac.uk} and Trevor J. Ponman \\
  School of Physics and Astronomy, University of
        Birmingham, Edgbaston, Birmingham B15 2TT, UK\\}
 \date{Accepted 2000 ??.
      Received 2000 ??;
      in original form 2000 ??}

\pagerange{\pageref{firstpage}--\pageref{lastpage}}
\pubyear{2000}

\begin{document}

\maketitle

\label{firstpage}

\begin{abstract}
  
  \noindent We compare the X-ray properties of loose and compact galaxy
  groups, using a combined sample of 42 groups. We find that we are unable
  to separate loose and compact groups on the luminosity-temperature
  relation, the luminosity-velocity dispersion relation or the velocity
  dispersion-temperature relation using equally weighted errors. This
  suggests that the distinction between compact and loose groups is not a
  fundamental one, and we argue that a more useful distinction is that
  between X-ray bright and X-ray faint systems.

  Given their similarity in X-ray properties, we combine the loose 
  and compact subsamples to derive relations based on the full sample.
  This provides the highest statistical quality results to date
  on the way in which the correlations in X-ray properties of low 
  mass systems depart from those seen in rich clusters.
  
\end{abstract}

\begin{keywords}
Intergalactic medium -- X-rays: galaxies
\end{keywords}


\section{Introduction}
\label{sec:intro}

Many galaxies in the local universe, including the Milky Way, are found in
dynamically bound groups. These groups have been divided into two broad
types: compact and loose. Compact groups are composed of galaxies separated
on the sky by only a few galactic radii, and as such they are more easily
identified than the more numerous loose groups, and hence have been
intensively studied.

The status of compact groups is still controversial. Such compact
configurations of galaxies, with modest velocity dispersions, would be
expected to result in galaxy merger rates greater than those observed
\cite{Zepf93}. Simulations of compact groups have shown that the predicted
merging rate is reduced if a significant amount of the group mass is
contained in a common halo (e.g.  \pcite{bode93}).  Moreover, compact cores
could be replenished by continuing infall of new galaxies
\cite{governato96}, or could be dense, bound configurations which form
temporarily within loose groups \cite{diaferio94}.  Alternatively, the low
merging rate would not pose a problem if compact groups were chance
alignments within loose groups \cite{mamon86} or filaments seen end on
\cite{hernquist95}, as in these cases the groups would not be as physically
compact as they appear.

Recent studies of compact groups which have probed further down the galaxy
luminosity function than the data from which the groups were originally
identified, indicate that many compact groups are located within overdense
environments (e.g.\pcite{decarvalho97,barton98,zabludoff98}).  On the basis
of galaxy distribution, there also seem to be different families of compact
groups, which may correspond to different dynamical stages of group
evolution \cite{ribeiro98}.

An alternative approach to investigation of the evolutionary status of
groups, is to study the X-ray emission from the intergalactic medium in
these systems. \scite{ponman96a} carried out an essentially complete survey
of the X-ray properties of the Hickson Compact Groups (HCGs), originally
identified by \scite{hickson82} -- the best studied sample of compact
groups. The results of this X-ray survey indicated that hot intergalactic
gas is found in association with at least 75\% of HCGs. This suggests that
most of these groups are real gravitationally bound systems.

Diffuse X-ray emission superficially similar to that seen in compact
groups, has also been noted in a number of loose groups
\cite{mulchaey96,mulchaey98b,helsdon99}.  If compact and loose groups are
truly different types of system, or if they represent very different stages
of group evolution, then one would expect to see differences in the
properties of the intragroup gas, and in particular in the correlations
involving X-ray luminosity and temperature, which reflect the relationship
between the gas, the potential well, and the galaxies it contains. In this
paper we combine the work of \scite{ponman96a} with the recent survey of
X-ray bright loose groups by \scite{helsdon99} to compare the X-ray
properties of loose and compact groups. Throughout this paper we take
H$_0$~=~50~km~s$^{-1}$Mpc$^{-1}$.


\section{The sample}
\label{sec:sample}

The properties of the compact groups used here are taken from an almost
complete survey of the redshift-accordant Hickson compact groups by
\scite{ponman96a}, while the properties of the loose groups are taken from
\scite{helsdon99}.  Detailed descriptions of the data reduction and
analysis may be found in these papers. \scite{ponman96a} used a combination
of {\it ROSAT} all-sky survey and pointed data. Point sources were
excluded, and a count rate and spectrum extracted within a radius
corresponding to 200kpc for all groups, except two in which a larger radius
of 500kpc was used. A hot plasma model was fitted to groups with pointed
data exceeding a 3$\sigma$ detection limit and luminosities derived using
the fitted model. For the RASS data a fixed spectral model was used to
derive a luminosity. This gave a total of 22 systems with detected diffuse
emission of which 16 had derived X-ray temperatures.

The groups examined in \scite{helsdon99} are based on pointed {\it ROSAT}
PSPC observations of 24 X-ray bright groups. These systems were originally
identified from three different sources, the optical group catalogues of
\scite{nolthenius93} and \scite{ledlow96} were examined to identify 15
X-ray bright groups, and then included were the 9 X-ray bright groups from
\scite{mulchaey98b}. \scite{helsdon99} excluded point sources and extracted
a count rate and spectrum within a radius determined for each group by
examining a smoothed image and group profile. A hot plasma model was then
used to obtain a temperature and derive a luminosity. This gave 24 groups
all with derived luminosities and temperatures.

Four systems (HCGs 42,62,68 and 90) are common to both samples. Despite the
somewhat different procedure used in the two studies, comparison of the
derived luminosities show that they typically agree to within 10\%, with
the temperatures agreeing to within approximately 15\%. These are compact
groups, and for our analysis below we use the parameters given in
\scite{ponman96a}, although for groups 42,62 and 90 the velocity
dispersions used are taken from \scite{zabludoff98}, as the larger number
of fainter galaxies they identify results in a more accurate velocity
dispersion. The full dataset is shown in Table~\ref{tab:data} and consists
of 20 loose groups and 22 compact groups, of which a subset of 16 compact
groups have temperature measurements.

It should be noted that neither the loose or compact group samples should
be regarded as being statistically complete in any way. However, we do not
believe that this will introduce any particular bias, other than the fact
that since we only use groups with detected diffuse X-ray emission, we do
not include systems with undetectably faint intergalactic gas. The two
samples sould rather be regarded as reasonably representative samples of
X-ray bright groups of each type.

\begin{table*}
  \center{\caption{\label{tab:data}Properties of the compact and loose groups}}
\begin{tabular}{lcccclccc}
\hline 
 & \multicolumn{2}{c}{Compact groups} & & & & \multicolumn{2}{c}{Loose groups} & \\
\cline{1-4} \cline{6-9}
Group name & Temperature   & log L  & velocity dispersion & & Group name & Temperature   & log L  & velocity dispersion \\  
 & (keV) & (erg s$^{-1}$) & km/s & & & (keV) & (erg s$^{-1}$) & km/s \\
\hline                                                                          
HCG12 & 0.89$\pm$0.12 & 42.31$\pm$0.08 & 269$\pm$99~ & & NGC315  & 0.85$\pm$0.07 & 42.15$\pm$0.15  & 122$\pm$43~\\     
HCG15 & 0.44$\pm$0.08 & 41.80$\pm$0.12 & 457$\pm$147 & & NGC383  & 1.53$\pm$0.07 & 43.31$\pm$0.02  & 466$\pm$48~\\     
HCG16 & 0.30$\pm$0.05 & 41.68$\pm$0.06 & 135$\pm$62~ & & NGC524  & 0.56$\pm$0.08 & 41.37$\pm$0.11  & 205$\pm$51~\\     
HCG33 & 0.61$\pm$0.30 & 41.77$\pm$0.11 & 174$\pm$80~ & & NGC533  & 1.06$\pm$0.04 & 42.95$\pm$0.02  & 464$\pm$55~\\     
HCG35 & 0.91$\pm$0.18 & 42.35$\pm$0.11 & 347$\pm$112 & & NGC741  & 1.08$\pm$0.06 & 42.66$\pm$0.03  & 434$\pm$48~\\     
HCG37 & 0.67$\pm$0.11 & 42.12$\pm$0.06 & 447$\pm$165 & & NGC1587 & 0.92$\pm$0.15 & 41.50$\pm$0.18  & 106$\pm$38~\\     
HCG42 & 0.82$\pm$0.03 & 42.16$\pm$0.02 & 211$\pm$36~ & & NGC2563 & 1.06$\pm$0.04 & 42.79$\pm$0.02  & 336$\pm$42~\\     
HCG48 & 1.09$\pm$0.21 & 41.58$\pm$0.14 & 355$\pm$212 & & NGC3607 & 0.41$\pm$0.04 & 41.59$\pm$0.03  & 421$\pm$172\\     
HCG51 & -             & 42.99$\pm$0.11 & 263$\pm$97~ & & NGC3665 & 0.45$\pm$0.11 & 41.36$\pm$0.10  & ~29$\pm$10~\\     
HCG57 & 0.82$\pm$0.27 & 41.98$\pm$0.21 & 282$\pm$84~ & & NGC4065 & 1.22$\pm$0.08 & 42.99$\pm$0.04  & 495$\pm$101\\     
HCG58 & 0.64$\pm$0.19 & 41.89$\pm$0.11 & 178$\pm$66~ & & NGC4073 & 1.59$\pm$0.06 & 43.70$\pm$0.01  & 607$\pm$94~\\  
HCG62 & 0.96$\pm$0.04 & 43.04$\pm$0.03 & 376$\pm$49~ & & NGC4261 & 0.94$\pm$0.03 & 42.32$\pm$0.02  & 465$\pm$57~\\     
HCG67 & 0.82$\pm$0.19 & 41.69$\pm$0.10 & 240$\pm$110 & & NGC4325 & 0.86$\pm$0.03 & 43.35$\pm$0.03  & 256$\pm$47~\\     
HCG68 & 0.54$\pm$0.15 & 41.27$\pm$0.26 & 170$\pm$63~ & & NGC4636 & 0.72$\pm$0.01 & 42.48$\pm$0.01  & 463$\pm$95~\\     
HCG73 & -             & 42.43$\pm$0.24 & ~95$\pm$57~ & & NGC5129 & 0.81$\pm$0.06 & 42.78$\pm$0.04  & 294$\pm$41~\\ 
HCG82 & -             & 42.29$\pm$0.14 & 708$\pm$326 & & NGC5171 & 1.05$\pm$0.11 & 42.92$\pm$0.05  & 424$\pm$84~\\     
HCG83 & -             & 42.81$\pm$0.12 & 501$\pm$185 & & NGC5846 & 0.70$\pm$0.02 & 42.36$\pm$0.02  & 368$\pm$67~\\ 
HCG85 & -             & 42.27$\pm$0.10 & 417$\pm$192 & & NGC6338 & 1.69$\pm$0.16 & 43.93$\pm$0.01  & 589$\pm$235\\     
HCG86 & -             & 42.32$\pm$0.14 & 302$\pm$139 & & NGC7619 & 1.00$\pm$0.03 & 42.62$\pm$0.02  & 253$\pm$96~\\
HCG90 & 0.68$\pm$0.12 & 41.48$\pm$0.09 & 193$\pm$35~ & & NGC7777 & 0.62$\pm$0.15 & 41.75$\pm$0.20  & 116$\pm$41~\\
HCG92 & 0.75$\pm$0.08 & 42.16$\pm$0.04 & 447$\pm$206 & &         &      &        &    \\
HCG97 & 0.87$\pm$0.05 & 42.78$\pm$0.02 & 407$\pm$150 & &         &      &        &    \\
\hline                                      
\end{tabular}
\end{table*}


\section{X-ray correlations}
\label{sec:results}

\begin{figure}
\psfig{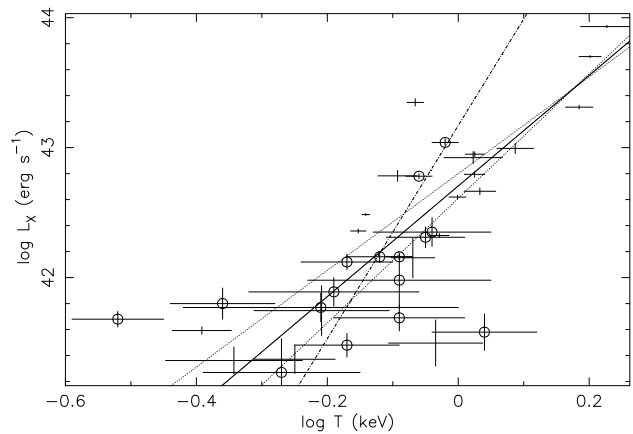}
\caption{\label{fig:ltrel}Relation between X-ray luminosity and temperature 
  for the combined compact (crosses with central circles) and loose (plain
  crosses) group samples. The bold line is best equally weighted fit to
  whole sample with 1 $\sigma$ error bounds marked with dotted lines.  The
  dash-dot-dash line is best fit to compact group sample using individual
  errors.}
\end{figure}

The relation between X-ray luminosity and temperature for loose and compact
groups is plotted in Figure~\ref{fig:ltrel}. It is clear that these two
parameters are significantly correlated (K=5.02, P$\ll$0.00001) across the
sample as a whole. In order to investigate whether there is any significant
difference between the relations for the loose and compact subsamples, we
perform a linear fit to the log$T$, log$L_X$ data to each subsample, and
derive confidence regions for these fits.  If these confidence regions are
disjoint then the two sets of groups have significantly different $L:T$
relations.

In calculating the best fit relations and their errors, the question arises
as to whether each data point should be weighted by its statistical error.
This is the correct approach provided that points deviate from the mean
relation only on account of these statistical errors. For the data shown in
Figure~\ref{fig:ltrel}, the variance about the mean trend is 6 times
greater than the variance expected from the statistical errors. This is not
surprising as it is clear from cluster studies that there is substantial
genuine scatter about the mean relationship (e.g.  \pcite{allen98}) which
is likely to be even more significant for groups (e.g.
\pcite{cavaliere97}). Under these circumstances, it is clearly not
appropriate to use weights based on the statistical error for each point,
and we adopt the alternative approach of weighting each point equally,
since the variance is dominated by real, rather than statistical, scatter.
In reality the variance about the mean trend will be a combination of both
real and statistical scatter. We therefore also include the results of
statistically weighted fits, for purposes of comparison.

For the equally weighted fits the $x$ and $y$ errors for each data point
(needed to provide the relative weights of offsets along each axis, and to
enable confidence intervals to be calculated) are calculated from the
observed scatter about the fitted trend. In order to determine the
appropriate weightings we initially fit a regression line to the entire
dataset using the \textsc{odrpack} package \cite{boggs90} which takes into
account the errors in both the $x$ and $y$ directions on each point. The
scatter about the best fit on both axes was then determined and used as an
estimate of the (equal) errors on each point. The data were then refitted
with a new regression line and the scatter was again determined.  This
iteration continued until a stable fit was found.

Table~\ref{tab:fits} shows the results of the different fits, using the
equally weighted errors, and for comparison, the fits obtained using
individual statistical errors on each point. The first column lists the
relation under consideration and whether each data point was weighted
equally or the individual errors on each point were used. The results of
fitting a straight line in log space to the compact, loose and combined
samples are then shown, along with 1$\sigma$ errors.  These errors were
derived using \textsc{odrpack}, and are based on the statistical scatter of
the points about the best fit line. As can be seen, in some cases there are
substantial differences in the slopes derived for the relations, depending
on how the data points are weighted. For example, the very steep $L:T$
slope (8.2) obtained by \scite{ponman96a} using statistical weighting, is
due in part to the strong leverage exerted by a small number of the
brightest groups, which have the smallest error bars. We believe that the
equally weighted results presented here give a more reliable measure of the
true correlations.

\begin{figure}
\hspace{1.25cm}
\psfig{file=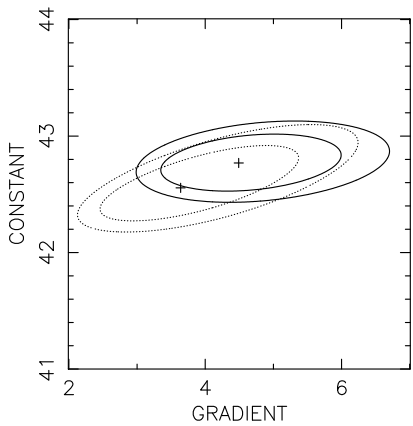,angle=0,width=6cm}
\caption{\label{fig:lt_chi}Confidence regions of straight lines fitted in 
  log space to the equally weighted L:T data of compact (dotted) and loose
  (solid) groups.  The crosses mark the best fit to each subset. Contours
  are 1 $\sigma$ and 90\% confidence.}
\end{figure}

Confidence regions in the slope:intercept plane have been calculated for
the loose and compact subsets, using the equally weighted data.  These
confidence regions are superposed in Figure~\ref{fig:lt_chi}. As can be
seen, there is considerable overlap, with the best fit for each sample
lying well within the 90\% confidence contour of the other, implying no
significant difference between the two datasets.  It therefore makes sense
to combine the two samples, and we plot in Figure~\ref{fig:ltrel} the best
fit line and one sigma error bounds for the entire sample with all points
weighted equally. Also plotted for comparison is the original relation
derived by \scite{ponman96a} for the compact group sample, in which the
errors on each individual point were used. This relation is clearly too
steep because two points (HCG62 and HCG97) have very small errors and
significantly steepen the line (c.f. the equally weighted compact group
sample in which the slope is 3.6). It should be noted that although the
slope of the fit to the compact group sample is much flatter using equally
weighted errors compared to using individual statistical errors, the lines
fitted to the compact and loose group samples are statistically equivalent
using either type of error (see Table~\ref{tab:fits}).

\begin{table*} 
\center{\caption{\label{tab:fits}Results of fits to the compact, loose and
    combined samples. Each fit is based on a straight line fit in log
    space. The best fits for the full dataset based on the equally weighted
    approach are marked in bold. $^a$ indicates all points were weighted
    equally, $^b$ indicates that individual errors on each point were
    used. Errors are 1$\sigma$. \dag This fit is noticeably affected by one
    discrepant point. Exclusion of this point steepens the slope to 2.9, as
    discussed in the main text.}}
\begin{tabular}{lccccccccc}
\hline 
 & & \multicolumn{2}{c}{compact groups} & & \multicolumn{2}{c}{loose groups} & & \multicolumn{2}{c}{full sample} \\
\cline{3-4} \cline{6-7} \cline{9-10}
relation & & gradient & intercept & & gradient & intercept & & gradient & intercept \\  
\hline                                                        
L:T$^a$        & &3.6$\pm$0.9&42.5$\pm$0.2  & &4.5$\pm$0.6&42.8$\pm$0.1  & &\bf{4.3$\pm$0.5}  &\bf{42.7$\pm$0.1}  \\     
L:T$^b$        & &8.2$\pm$2.7&43.17$\pm$0.26& &4.5$\pm$0.8&42.95$\pm$0.09& &5.0$\pm$0.7  &43.01$\pm$0.07\\     
L:$\sigma$$^a$ & &2.3$\pm$0.6&36.5$\pm$1.6  & &2.3$\pm$0.5&36.8$\pm$1.1  & &~~\bf{2.4$\pm$0.4} \dag  &\bf{36.4$\pm$1.0}  \\
L:$\sigma$$^b$ & &4.9$\pm$2.1&30.0$\pm$5.1  & &4.6$\pm$1.4&31.0$\pm$3.6  & &4.7$\pm$0.9  &30.6$\pm$2.3  \\     
$\sigma$:T$^a$ & &1.4$\pm$0.3&2.64$\pm$0.06 & &1.9$\pm$0.5&2.55$\pm$0.08 & &\bf{1.7$\pm$0.3}  &\bf{2.60$\pm$0.05} \\     
$\sigma$:T$^b$ & &0.9$\pm$0.3&2.55$\pm$0.05 & &1.1$\pm$0.3&2.57$\pm$0.03 & &1.0$\pm$0.2&2.57$\pm$0.03 \\
\hline                                      
\end{tabular}
\end{table*}

\begin{figure}
\hspace{1.25cm}
\psfig{file=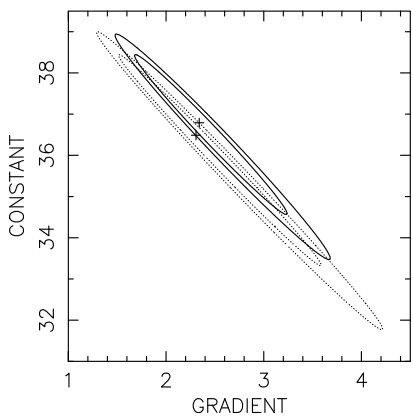,angle=0,width=6cm}
\caption{\label{fig:lv_chi}Confidence regions of straight lines fitted in 
  log space to the equally weighted L:$\sigma$ data of compact (dotted) and
  loose (solid) groups. The crosses mark the best fit to each subset.
  Contours are 1 $\sigma$ and 90\% confidence.}
\end{figure}

The relationship between X-ray luminosity and group velocity dispersion is
also examined in an identical way to the $L:T$ relation.
Table~\ref{tab:fits} shows the results of the different fits to the whole
$L:\sigma$ dataset and the separate loose and compact subsamples. Maps of
$\chi^{2}$ were produced for the two subsamples and once again a clear
overlap is seen in the confidence regions (Figure~\ref{fig:lv_chi}).

\begin{figure}
\psfig{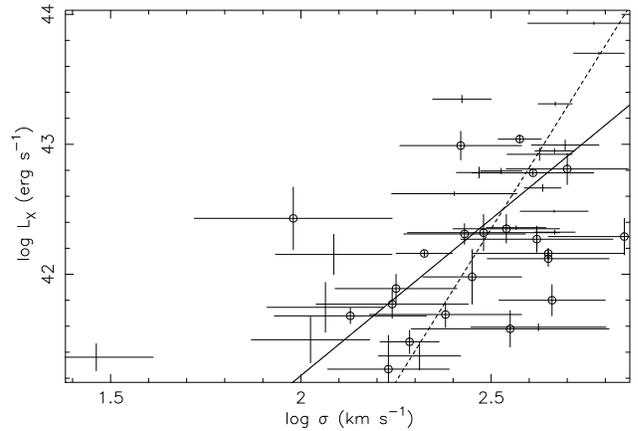}
\caption{\label{fig:lvrel}Relation between X-ray luminosity and velocity 
  dispersion. Data symbols are same as in Figure~\ref{fig:ltrel}. The solid
  line shows best fit equally weighted line to whole sample and the dashed
  line shows individually weighted fit to whole sample.}
\end{figure}

The relationship between X-ray luminosity and velocity dispersion for the
full sample is plotted in Figure~\ref{fig:lvrel}. The correlation between
these parameters is strong (K=3.92, P=0.0001), and also shown are the best
fit individually weighted (dashed line) and equally weighted (solid line)
fits to the data. As can be seen there is a very significant difference
between the two lines. There is one point with a very low velocity
dispersion (NGC 3665, 29~km~s${-1}$) which stands out on this graph. A
virialised system must have a minimum mean density related to the density
of the universe which can be used to constrain the velocity dispersion of a
virialised group to be greater than 100~km~s${-1}$ \cite{mamon94}. A
velocity dispersion of 29~km~s${-1}$ clearly lies below this limit. Given
that the X-ray emission suggests a collapsed system, the real velocity
dispersion of the system is likely to be much higher than has been measured
from the four catalogued members. Exclusion of this point has a significant
effect on the gradient of the equally weighted line, increasing the slope
to 2.9 which is within 2$\sigma$ of the individually weighted line.

\begin{figure}
\psfig{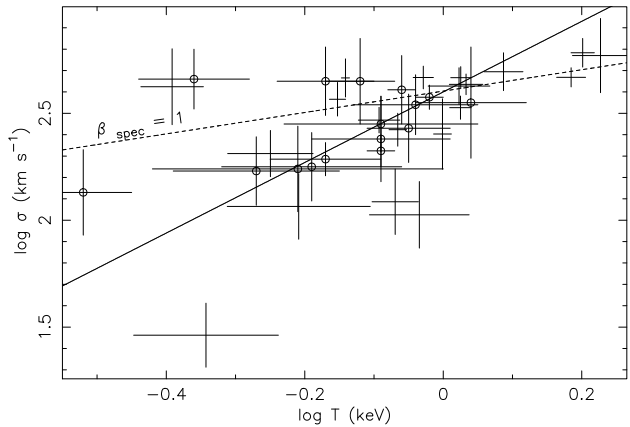}
\caption{\label{fig:vtrel}Relation between velocity dispersion and 
  temperature. Data symbols are same as in Figure~\ref{fig:ltrel}. The
  solid line shows best fit equally weighted line to whole sample and the
  dashed line shows the line $\beta_{spec}=1$.}
\end{figure}

\begin{figure}
\hspace{1.25cm}
\psfig{file=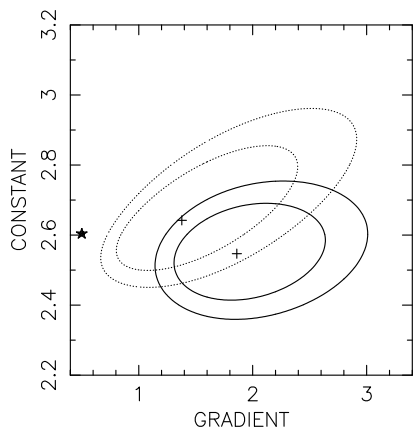,angle=0,width=6cm}
\caption{\label{fig:vt_chi}Confidence regions of straight lines fitted in 
  log space to the equally weighted $\sigma$:T data of compact (dotted) and
  loose (solid) groups.  The crosses mark the best fit to each subset. The
  solid star marks the position of $\beta_{spec}=1$.  Contours are 1
  $\sigma$ and 90\% confidence.}
\end{figure}

The relationship between velocity dispersion and temperature is shown in
Figure~\ref{fig:vtrel}. Once again a series of fits are carried out on the
samples (Table~\ref{tab:fits}) and the $\chi^{2}$ maps
(Figure~\ref{fig:vt_chi}) show significant overlap. Also plotted on
Figure~\ref{fig:vtrel} is the equally weighted regression line and the line
for $\beta_{spec}=1$, where $\beta_{spec}$ is the ratio of the specific
energy in the galaxies to that in the gas. The locus $\beta_{spec}=1$ is
plotted on Figure~\ref{fig:vt_chi} and clearly shows that $\beta_{spec}=1$
is a poor fit to both subgroups of data.

\section{Discussion and Conclusions}
\label{sec:dis}

As can be seen from the $\chi^{2}$ maps for each of the three relations,
the compact and loose group data do not appear to differ significantly. The
best fit for each subset (compact/loose) always lies within the 90\%
confidence region of the other subset, and there is much overlap of the 1
sigma confidence regions.  Although there are some differences in the
regression lines obtained when using statistically or equally weighted
errors, the compact and loose group samples are still equivalent to one
another when using either type of error (see Table~2). The X-ray surface
brightness profiles of compact galaxy groups (e.g.
\pcite{ponman93a,david95,helsdon99,lloyddavies99}) are also comparable with
those derived for loose groups \cite{helsdon99} and are flatter than those
observed in galaxy clusters, providing further evidence that compact and
loose groups are similar systems.

Given this similarity, the results shown in bold in Table~\ref{tab:fits},
derived from the full sample of 42 groups, represent the best estimates
available of these relations for galaxy groups. Though, as discussed in
section~3, the slope of the $L:\sigma$ relation given in the Table is
biased low by one outlying system which probably has an incorrect velocity
dispersion. These results are derived using the equally weighted errors
which should be more reliable than using individual statistical errors,
given the observed scatter in the relationships.

The steepening of the $L:T$ and $\sigma:T$ relations in low temperature
systems support preheating models (see \pcite{helsdon99} for more details),
in which injection of energy from the epoch of galaxy formation lowers the
density and raises the temperature of the intergalactic gas in galaxy
groups. Models of this process \cite{cavaliere99,balogh99} predict that the
mass-temperature relation should rapidly steepen at a temperature related
to the temperature to which the gas was preheated. As $M\propto\sigma^2$ we
should also see this effect as a steepening of the $\sigma:T$ relation, as
is observed below $T=1$~keV in Fig.5. Also noticeable is the large scatter
about the mean trends in all three of the correlations examined here --
especially in the poorest systems. This finds a natural interpretation in
terms of the variation in detailed evolutionary history in such sparse
systems, between one group and another.

What implications for the relationship between compact and loose groups
follow from the apparent indistinguishability in their X-ray properties?
One possibility is clearly that compact groups are simply fortuitous
alignments of galaxies within loose groups \cite{mamon86} -- however the
accumulation of data showing clear evidence for strong galaxy interactions
within HCGs has convinced even most sceptics that many of them are
genuinely dense in three dimensions \cite{mamon99}. On the other hand, the
results presented above, in conjunction with deep optical surveys showing
much more extended distributions of galaxies associated with compact groups
\cite{decarvalho97,barton98,zabludoff98}, argue that they are not
fundamentally different from loose groups.

This apparently leaves two viable models for the nature of compact groups.
In the model of \scite{diaferio94}, compact groups are temporary bound
configurations which form within loose groups, whilst \scite{governato96}
propose a model in which compact cores are regenerated by continuing infall
of galaxies, whilst merger rates are reduced by halo stripping of the group
members. The model of \scite{diaferio94} appears to conflict with the
observed X-ray properties of compact groups. In the simulations presented
by these authors, compact groups are produced at a variety of locations
within their natal loose group, during its evolution. The majority of these
temporary bound groupings do not occur at the centre of the main group. In
this case, one would expect that compact groups would frequently be offset
from the X-ray centroid associated with the core of the larger potential
well of the group as a whole. This is not what is observed. X-ray studies
show that HCGs associated with X-ray emission are invariably at the core of
that emission.

The model of \scite{governato96} appears to account well for the observed
properties of X-ray bright compact groups, and for the fact that they are
embedded within looser overdense configurations. The fact that such X-ray
bright systems invariably contain a fairly bright early type galaxy near
their centre is explained in this model as the result of early merging
activity in the collapsing group core.  The existence of X-ray bright loose
groups with essentially identical X-ray properties follows immediately from
the fact that such systems only satisfy the isolation and compactness
criteria required of compact groups for a fraction of the time.

There is, however, a second type of compact group, which cannot be
accounted for by the model of \scite{governato96}.  These groups are not
dominated by an early type galaxy, they may show signs of galaxy
interactions in the optical \cite{deoliveira94} or HI
\cite{verdesmontenogro99}, but they show either weak or no diffuse X-ray
emission ($L_X < 10^{42}$ erg s$^{-1}$), and they typically have lower
velocity dispersions that the X-ray bright groups. \scite{ribeiro98}
suggest that such systems (their ``core+halo'' and ``compact'' classes) may
be in a more advanced evolutionary state than the systems which represent
the cores of more extensive loose groups. However, the common association
of extensive hot gas halos with the latter class of groups shows that this
cannot be the case. There is no way that systems like HCG42, HCG62 and
HCG97 can lose their X-ray bright halos as their evolution proceeds, so
they cannot evolve into X-ray faint systems like HCG16. Also the fraction
of ellipticals and velocity dispersion are typically higher in the X-ray
bright systems, and these are most unlikely to drop as the system evolves.

It seems that either the X-ray faint groups (both loose and compact) are an
{\it earlier} evolutionary stage, or they are simply different from the
X-ray bright systems. Their low X-ray luminosity suggests that either these
systems have, like the Local Group, not yet collapsed (so that their
intergalactic gas has not been compressed to the point where it emits
detectable X-ray emission), or that their potential wells are too shallow
to concentrate preheated gas sufficiently to achieve a detectable X-ray
surface brightness. In the case of compact groups in which strong galaxy
interactions are seen, it is clear that the core of the system must have
collapsed to a dense state, so the second effect must be the dominant one.



\section{Acknowledgements}
We thank Edward Lloyd-Davies for useful comments and advice, and the
referee for suggesting improvements to the paper. SFH acknowledges
financial support from the University of Birmingham. This work made use of
the Starlink facilities at Birmingham.


\bibliography{../reffile}
\label{lastpage}

\end{document}